\documentclass[preprint,amsmath]{revtex4-1}

\usepackage{blindtext}
\usepackage{epsfig}
\usepackage{amsmath}
\usepackage{amssymb}
\usepackage{color}
\usepackage[normalem]{ulem}
\usepackage{caption}
\usepackage{subcaption}
\usepackage{mathtools}
\usepackage{color}
\usepackage{epstopdf}
\usepackage{array}
\usepackage{wasysym}

\begin{document}
\title{Limit Forms of the Distribution of the Number of Renewals}

\author{Stanislav Burov }
\email{stasbur@gmail.com}

\affiliation{Physics Department, Bar-Ilan University, Ramat Gan 5290002,
Israel}

\pacs{PACS}

\begin{abstract}
 In this work the asymptotic properties of  $Q_t(N)$ ,the probability of the number of renewals ($N$), that occur during time $t$ are explored. While the forms of the distribution at very long times, i.e. $t\to\infty$, are very well known and are related to the Gaussian Central Limit Theorem or the L\'{e}vy stable laws, the alternative limit of large number of renewals, i.e. $N\to\infty$, is much less noted. We address this limit of large $N$  and find that it attains a universal form that solely depends on the analytic properties of the distribution of renewal times. 
 Explicit formulas for $Q_t(N)$ are provided, together with corrections for finite $N$ and the necessary conditions for convergence to the universal asymptotic limit. Our results show that the Large Deviations rate function for $N/t$ exists and attains an universal linear growth (up to logarithmic corrections) in the $N/t\to\infty$ limit. This result holds irrespective of the existence of mean renewal time or presence of power-law statistics.
 

\end{abstract}

\maketitle

\section{Introduction}
\label{introductSec}

An important simplification in statistical description of many processes in Physics is the assumption of renewal. 
This assumption states that a time dependent process is simply renewed/reset at specific point in time and its behavior is independent from the history that precedes this time-point~\cite{CoxBook}. The times between renewals can depend on the development of the process but are independent one from each other. 
Examples include, but are not limited to, magnetic spin fluctuations~\cite{Godreche}, continuous time random walk  in complex media~\cite{ScherMontrol} and models of cell growth and division~\cite{Amir2014}. Recently the concept of stochastic resetting became very popular~\cite{EvansMajumdar2011,Reuveni2016,Chechkin2018,EvanMajumdarScher2020}. In this scheme the renewal is not assumed as the original part of the process but rather imposed externally. It was shown that different protocols of resetting can drive the process to  non-equilibrium steady states that differ from the, some-times non-existent, steady states of the explored system. 

It is often of great interest to acquire the knowledge of the number of renewals that occurred during time $t$. 
A classical example is the number of buses that arrived to a bus-station during a given period of time, if the assumption of independence of time-intervals between arrivals is approximately correct. 
For transport in random media the framework of continuous time random walk (CTRW) describes the motion of a particle as a series of jumps between spatially scattered traps and random waiting times at each trap ~\cite{Bouchaud,KlafterMetzler}. The number of renewals for CTRW is exactly the number of jumps that have been performed.
By conditioning on the number of renewals it is possible to calculate the positional probability density of the process and various other quantities~\cite{Sokolov08}. This method of conditioning is termed subordination~\cite{Bouchaud,Fogedby1994,Lindenberg2005,Gorska2010,Magziarz2007,Magdziarz2008,Eule_2009} and it can even be generalized to  cases of quenched disorder where the CTRW framework doesn't hold~\cite{BurovBarkai,Burov2017,Burov2020Quenched}. 
Interestingly, it was found that the positional probability density for CTRW can display universal non-Gaussian, but rather exponential, behavior~\cite{BarkaiBurov2020,Wang2020} that is experimentally observed in different complex systems, such as glasses~\cite{Kob01}, colloids~\cite{WeitzGlass,Kilfoil}, molecular motion on a solid-liquid interface~\cite{Shwartz13,Shwartz17}, active gels~\cite{MIzuno} and many other disordered systems~\cite{Xue2020Diffusion,Shin2019Anomalous,Roichman2020,Witzel2019Heterogeneities,Cherstvy2019Non,Mejia2020Tracer} (See~\cite{Granick02} and~\cite{ChechkinPRX} for additional discussion). The probability to observe large number of renewals for a given time $t$ plays a keynote role in determining this finding~\cite{BarkaiBurov2020}. So while it might be not that fascinating to calculate the probability of a large number of arrived buses, the probability of a large number of renewals is of great importance for studying the properties the transport in complex systems and it is the focus of this manuscript.   

The manuscript is organized as follows. In Section~\ref{renewsec} we define the renewal process and our quantity of interest, i.e. the number of renewals. We briefly restate the known results for the distributions of this quantity in the $t\to\infty$ limit and summarize them in Table~\ref{table:1}. Section~\ref{qtnSec} is devoted to the probability to observe a large number of renewals in a finite time. The asymptotic form of this probability is describe by Eq.~\eqref{qtnassymptotic} and the explicit proof is supplied in Section~\ref{qtnproofSec}. The development of corrections to this formula for finite number of renewals is the purpose of Section~\ref{sechighorderpsi}. In Section~\ref{seclargedev} we tie together the limits of large number of renewals and large measurement time by using the approach of the theory of Large Deviations. We conclude with a summary. The two appendixes supply the necessary information about hyper-geometric functions, that are exploited for development of our main results.

\section{The Renewal Process}
\label{renewsec}

The definition of the renewal process is based on existence of a positively defined random variable $\tau$ with the probability density function (PDF) provided by $\psi(\tau)$. $\tau$ can describe the failure time of an electric component, a sojourn time of a particle in specific area or the time between the arrival of two successive buses. For the example of failure time, the process starts with a new component at time $t=0$. The component fails at time $\tau_1$ and it is immediately replaced by a new component. The failure time of the second component is $\tau_2$ and it is replaced (again) by a third component. The components are constantly replaced, and the count down till new replacement immediately starts, hence the name "renewal process". The failure of the $N$th component is $\tau_N$ and the total time till replacement of this component, $t_N$, is
\begin{equation}
t_N=\sum_{i=1}^N\tau_i.
    \label{timedef_n}
\end{equation}
All the different $\tau_i$ are IID random variables. $t_N$ is a random variable and for given $\psi(\tau)$ one can calculate its PDF $S_N(t_N)$. While for $t_N$ in Eq.~\eqref{timedef_n} the number of renewals, $N$, is fixed, one can fix the measurement time $t$ and ask what is the number of renewals, $N_t$ that occurred during this time. Specifically,
\begin{equation}
    t=\sum_{i=0}^{N_t}\tau_i+u_t,
    \label{numberofrenewdef}
\end{equation}
where $u_t$ is the backward recurrence time that is present since $t$ is set up before the $(N_t+1)$th renewal takes place, i.e.   $0<u_t<\tau_{N_t+1}$.
When $t$ is fixed, $N_t$ is the random variable that is distributed according to $Q_t(N_t)$. In the following we are interested in the form of $Q_t(N_t)$ in the limits of large $t$ and large $N_t$. 

The two random variables $t_N$ and $N_t$ are connected to each other and so are the functions $Q_t(N_t)$ and $S_N(t_N)$. As pointed out in~\cite{CoxBook}, $N_t< N$ if and only if $t_N>t$. Indeed, when the time to accomplish exactly $N$ renewals is larger than some preset time $t$, it means that  if one takes the same $\tau_i$s (that were used to construct $t_N$) and use them to construct $t$ according to Eq.~\eqref{numberofrenewdef}, then $N_t$ can't be bigger than $N$. On the other hand if $N\geq N_t+1$ then $t_N>t$, due to the fact that $0<u_t<\tau_{N_t+1}$. 
Then we can write that $\theta\left(N-N_t\right)=1-\theta\left(t-t_N\right)$, where $\theta(\dots)$ is the Heaviside step function. Averaging over all possible realizations of $\tau_i$s we obtain that 
\begin{equation}
\langle \theta\left(N-N_t\right) \rangle = 
1 - \langle \theta\left(t-t_N\right) \rangle.
    \label{thetarelation}
\end{equation}
Since $\langle \theta\left(N-N_t\right) \rangle - \langle \theta\left(N+1-N_t\right) \rangle = Q_t(N) $ and $1-\langle \theta\left(t-t_N\right) \rangle=\int_0^t S_N(t')dt'$, we obtain from Eq.~\eqref{thetarelation} that
\begin{equation}
Q_t(N)=\int_0^t \left[S_N(t')-S_{N+1}(t')\right]dt'.
    \label{probrelations01}
\end{equation}
In Eq.~\eqref{probrelations01} we omitted the explicit $N_t$ and $t_N$ form in $Q_t(N_t=N)$ and $S_N(t_N=t)$. Equations \eqref{thetarelation} and \eqref{probrelations01} relate the probability to observe $N$ renewals (that occur during  fixed time $t$)  to the PDF of a random $t$ that is the sum of $N$ (fixed) random variables. 
Due to the fact that $S_N(t)$ is a sum of IID random variables we apply the Laplace transform on both sides of Eq.~\eqref{probrelations01} and obtain
\begin{equation}
{\hat Q}_s(N) = \frac{1-{\hat \psi}(s)}{s} {\hat \psi}(s)^N ,
 \label{probrelLaplace}   
\end{equation}
where ${\hat Q}_s(N)=\int_0^\infty Q_t(N)\exp(-st)\,dt$ and ${\hat \psi}(s)=\int_0^\infty \psi(\tau)\exp(-s\tau)\,d\tau$.
For any given $\psi(\tau)$ Eq.~\eqref{probrelLaplace} provides the appropriate representation of $Q_t(N)$ in Laplace space and the only thing that is left in order to obtain an expression for $Q_t(N)$, is to properly take the inverse Laplace transform.


 In the limit of $t\to\infty$ the behavior of $Q_t(N)$ is dictated by the central limit theorem (CLT) or L\'{e}vy stable laws and is separated into three general classes. Only the behavior of $\psi(\tau)$ for large $\tau$ is important. While all the results for $t\to\infty$ are fairly known, we present the results in Table~\ref{table:1} and shortly describe the way of their derivation for completeness of presentation.

\begin{table}
\begin{center}
 \begin{tabular}{|| m{4em}|| m{15em} | m{19em} ||} 
 \hline
 Case \# 
 &
 Properties of $\psi(\tau)$
 &  $Q_t(N)\quad _{(t\to\infty)}$  \\ [0.5ex] 
 \hline\hline
 \begin{center}I\end{center} &
\begin{center}Finite mean $\mu$ and variance $\sigma^2$ \end{center} & \vspace{0.0cm}\begin{center} $\displaystyle{\mu^{3/2}\exp\left({-\frac{(N-t/\mu)^2}{2\sigma^2 t/\mu^3}}\right)\Big/{\sqrt{2\pi\sigma^2 t}}}$\vspace{0.00cm}\end{center}  \\ 
 \hline
  \begin{center}II\vspace{0.5cm}\end{center} &
 Diverging variance and finite mean 
 \begin{center}
    $\displaystyle{\psi(\tau)\underset{t\to\infty}{\sim}A\tau^{-1-\alpha}/\Gamma(-\alpha)}$
 {$\qquad(1<\alpha<2)$} 
 \end{center} & 
 \begin{center}
 $ \displaystyle{\mu^{1+1/\alpha}l_{\alpha,1}\left(\frac{t/\mu-N}{\left(At\right)^{1/\alpha}\big/\mu^{1+1/\alpha}}\right) \Big/
 {\left(At\right)^{1/\alpha}}
 }$
 \vspace{0.5cm}
 \end{center}
 \\
 \hline
  \begin{center}III\vspace{0.5cm}\end{center} &
 \begin{center}
 Diverging variance and  mean 
     $\displaystyle{\psi(\tau)\underset{t\to\infty}{\sim}A\tau^{-1-\alpha}/|\Gamma(-\alpha)|}$
 {$\qquad(0<\alpha<1)$} 
 \end{center}
 & 
 \begin{center}
 $\displaystyle{\frac{t}{\alpha A^{1/\alpha} } N^{-1-1/\alpha} l_{\alpha,1}\left(\frac{t}{\left(AN\right)^{1/\alpha}}\right)}$
 \end{center}
 \vspace{0.2cm}
 \\ [1ex] 
 \hline
\end{tabular}
\end{center}
\caption{Limit cases of observing $N$ renewals during time $t$, i.e. $Q_t(N)$,  for large measurement times. The formulas are developed in Sec.~\ref{secgausslimit}, Sec.~\ref{secstablelaw01} and Sec.~\ref{secstablelaw02}. For cases \#I and \#II the typical $N$ is of the order of $t$ while in case \#III $N\sim t^\alpha$. When $N>>t$ this description fails and new form (Eq.~\eqref{qtnassymptotic}) emerges.}
\label{table:1}
\end{table}

\subsection{Gaussian  Limit}
\label{secgausslimit}
The first case we address is the case when the properties of $\psi(\tau)$ force $t_N$ to be in a basin of attraction of the Gaussian behavior. 
This means that when both the first moment $\mu=\int_0^\infty \tau\psi(\tau)\,d\tau$ and the variance $\sigma^2=\int_0^\infty \tau^2\psi(\tau)\,d\tau-\mu^2$ are both finite, then according to the CLT  $S_N(t)$ is 
\begin{equation}
S_N(t)\underset{N\to\infty}{\sim}\frac{1}{\sqrt{2\pi \sigma^2 N}}e^{-\frac{\left(t-\mu N\right)^2}{2\sigma^2 N}}.
    \label{gaussianpdf01}
\end{equation}
In the language of Eq.~\eqref{thetarelation} the CLT states that 
\begin{equation}
\Big\langle \theta(y-[t_N-\mu N]/\sigma \sqrt{N})\Big\rangle
\underset{N\to\infty}{\longrightarrow} \int_{-\infty}^y\exp(-u^2/2)\,du/\sqrt{2\pi},
\label{gaussslimittheta}
\end{equation}
this means that  we can write $t=\mu N+y\sigma \sqrt{N}$ and $y$ has a normal distribution with zero mean and unit variance. This change of variables $t\leftrightarrow y$ and the statement of Eq.~\eqref{gaussslimittheta} is utilized for finding the behavior of $Q_t(N)$. We start with a change of variables of the form $N=a t +b y\sqrt{t} $ where $a$,$b$ are positive constants that will be exactly found below and $y$ is the new variable. Then we write $\theta(t-t_N)$ as $\theta([t-N/a-(t_N-N/a)]/\sigma\sqrt{N})$. Since $(t-N/a)/\sigma\sqrt{N}=-by/a\sigma\sqrt{a+by/\sqrt{t}}$ we choose $a=1/\mu$ and $b=\sigma/\mu^{3/2}$ and take the $t\to\infty$ limit to obtain
\begin{equation}
\theta\left(t-t_N\right)\underset{t\to\infty}{\longrightarrow}
\theta\left(-y-[t_N-\mu N]/\sigma\sqrt{N}\right).
    \label{thetaNtlimit}
\end{equation}
When $t$ is diverging so is $N$, then according to Eq.~\eqref{gaussslimittheta} and  Eq.~\eqref{thetarelation}
\begin{equation}
\Big{\langle} \theta\left(y-\frac{N_t-t/\mu}{\sigma\sqrt{t}/\mu^{3/2}}\right)\Big{\rangle}\underset{t\to\infty}{\longrightarrow}\frac{1}{\sqrt{2\pi}}\int_{-\infty}^y \exp(-u^2)\,du. 
    \label{qtNgausscumul}
\end{equation}
This means that that the distribution of $N_t$ is Gaussian with mean $t/\mu$ and variance $t\sigma^2/\mu^3$, i.e. 
\begin{equation}
Q_t(N)\underset{t\to\infty}{\sim}\frac{1}{\sqrt{2\pi\sigma^2 t/\mu^3}}e^{-\frac{(N-t/\mu)^2}{2\sigma^2 t/\mu^3}}.
    \label{qtngaussianfinal}
\end{equation}

\subsection{Effect of power-laws}

When the behavior of $\psi(\tau)$ for large $\tau$ is of the form
\begin{equation}
    \psi(\tau)\underset{\tau\to\infty}{\sim}\tau^{-1-\alpha} \qquad \left(0<\alpha<2\right).
    \label{powerlawpsi}
\end{equation}
two distinct casses are present. 
When $1<\alpha<2$ the mean renewal time diverges while the variance is finite and when $0<\alpha<1$ also the mean $\tau$ diverges. These two cases define two different forms of L\'{e}vy stable laws for $S_N(t)$ and then accordingly produce two different forms of $Q_t(N)$.

\subsubsection{$0<\alpha<1$}
\label{secstablelaw01}
For this case the asymptotic form of $\psi(\tau)$ in the limit of $\tau\to\infty$ is $A\tau^{-1-\alpha}\Big/|\Gamma(-\alpha)|$. The L\'{e}vy stable law takes the form
\begin{equation}
   \Big\langle \theta\left(\eta-\frac{t_N}{(A N)^{1/\alpha}}\right)\Big\rangle
   \underset{N\to\infty}{\longrightarrow}\int_0^\eta l_{\alpha,1}(u)\,du,
   \label{stablelevyalp01}
\end{equation}
where $l_{\alpha,1}(u)$ is one-sided L\'{e}vy PDF whose Laplace pair is $\exp(-s^\alpha)$. 
In order to find $Q_t(N)$ we make a change of variables of the form $N=t^\alpha/(Az^\alpha)$. Then using the fact that $\theta(t-t_N)=\theta(\frac{t-t_N}{(AN)^{1/\alpha}})$ and that $N$ is diverging when $t\to\infty$, we obtain from Eqs.~\eqref{thetarelation} and~\eqref{stablelevyalp01} 
\begin{equation}
\Big\langle\theta\left(N-N_t\right)\Big\rangle
\underset{t\to\infty}{\longrightarrow}\int_z^\infty l_{\alpha,1}(u)\,du.
    \label{thetalevyalp01}
\end{equation}
Since $\langle\theta(N-N_t)\rangle$ is cumulative probability we obtain that
\begin{equation}
Q_t(N)\underset{t\to\infty}{\sim} \frac{t}{\alpha A^{1/\alpha} } N^{-1-1/\alpha} l_{\alpha,1}\left(\frac{t}{\left(AN\right)^{1/\alpha}}\right)
\qquad\left(0<\alpha<1\right)
    \label{poweralp01qtn}
\end{equation}

\subsubsection{$1<\alpha<2$}
\label{secstablelaw02}

In this case $\psi(\tau)$ still has the asymptotic form $A\tau^{-1-\alpha}/\Gamma(-\alpha)$ when $\tau\to\infty$, but since $1<\alpha<2$ the first moment $\mu=\int_0^\infty \tau\psi(\tau)\,d\tau$ is finite. The L\'{e}vy stable law takes the form 
\begin{equation}
   \Big\langle \theta\left(\eta-\frac{t_N-\mu N}{(A N)^{1/\alpha}}\right)\Big\rangle
   \underset{N\to\infty}{\longrightarrow}\int_{-\infty}^\eta l_{\alpha,1}(u)\,du,
   \label{stablelevyalp02}
\end{equation}
where $l_{\alpha,1}(u)$ is the asymmetrical L\'{e}vy PDF that have a Laplace pair $\exp(-s^\alpha)$ and for $1<\alpha<2$ is defined for $-\infty<u<\infty$. For this case the change of variables is $N=t/\mu+z\left( A t\right)^{1/\alpha} /\mu^{1+1/\alpha}$ and accordingly $\theta(t-t_N)=\theta(-\mu^{1/\alpha}z(1/\mu+(A t)^{1/\alpha}z/t\mu^{1+1/\alpha})^{-1/\alpha}-\frac{t_N-\mu N}{(AN)^{1/\alpha}})$ that in the $t\to\infty$ limit takes the form $\theta(-z-\frac{t_N-\mu N}{(AN)^{1/\alpha}})$. Then Eq.~\eqref{stablelevyalp02} and Eq.~\eqref{thetarelation} yield
\begin{equation}
\Big\langle\theta\left(N-N_t\right)\Big\rangle
\underset{t\to\infty}{\longrightarrow}\int_{-z}^\infty l_{\alpha,1}(u)\,du.
    \label{thetalevyalp02}
\end{equation}
Since $\langle\theta(N-N_t)\rangle$ is cumulative probability we obtain that
\begin{equation}
Q_t(N)\underset{t\to\infty}{\sim} \frac{1}{A^{1/\alpha}t^{1/\alpha}\big/\mu^{1+1/\alpha}} l_{\alpha,1}\left(\frac{t/\mu-N}{A^{1/\alpha}t^{1/\alpha}\big/\mu^{1+1/\alpha}}\right)
\qquad\left(1<\alpha<2\right)
    \label{poweralp02qtn}
\end{equation}

\section{The probability to observe a very large number of renewals in a finite time.}
\label{qtnSec}

The probability $Q_t(N)$ was calculated in Sec.~\ref{renewsec} in the limit of long times. For the cases when mean renewal time, $\langle\tau\rangle=\mu$, was finite the calculated formula described deviations around the average number of renewals, i.e., $\langle N\rangle=t/\mu$. The case when the time $t$ can be large but finite, while the interest is in the limit when $N\to\infty$ is quite different from what is described in Table~\ref{table:1}. The derivations of Sec.~\ref{renewsec} rely on the CLT or on the stable-laws that do not take into account atypical behavior of extreme events. In order to observe large number of renewals in a given $t$, all of those renewals must prolong for a very short time. This means that $Q_t(N)$ in this limit should depend solely on the short $\tau$ properties of $\psi(\tau)$, while those properties are not part of the details of Table~\ref{table:1}.

 We assume that  $\psi(\tau)$ is analytic in the $\tau\to 0$ limit and its Taylor expansion takes the form 
\begin{equation}
\psi(\tau)\underset{\tau\to 0}{\sim}\sum_{i=0}^\infty C_{A+i}\tau^{A+i},
    \label{taylorexpansion}
\end{equation}
where $A\geq 0$ is an integer that describes the leading order of the Taylor expansion. The next step is to exploit Eq.~\eqref{probrelLaplace} and use the large $s$ expansion of $\psi(s)$ that provides the $t\to 0$ limit, as opposed to the $t\to\infty$ limit explored in Sec.~\ref{renewsec}. 
Eq.~(\ref{taylorexpansion}), and Tauberian Theorem~\cite{Weiss}, dictates the form of ${\hat{\psi}}(s)$ in the limit $s\to\infty$
\begin{equation}
    {\hat{\psi}}(s)\underset{s\to\infty}{\sim}\sum_{i=0}^\infty\frac{C_{A+i}\Gamma\left(A+i+1\right)}{s^{A+i+1}}.
    \label{tauberianpsi}
\end{equation}
By introducing Eq.~(\ref{tauberianpsi}) into Eq.~\eqref{probrelLaplace} we obtain the $s\to\infty$ expansion of ${\hat{Q}}_s(N)$
\begin{equation}
{\hat{Q}}_s(N)\sim\frac{\left[C_A\Gamma(A+1)\right]^N}{s^{N(A+1)+1}}\left(1+
\sum_{i=1}^\infty\frac{C_{A+i}\Gamma(A+i+1)}{C_A \Gamma(A+1) s^{i}}\right)^N
\left(1-\sum_{i=0}^\infty\frac{C_{A+i}\Gamma(A+i+1)}{s^{A+i+1}}\right).
    \label{qsnexp02}
\end{equation}
On the right hand side of Eq.~\eqref{qsnexp02} we take the largest power of $s$ and by application of binomial expansion obtain
\begin{equation}
{\hat{Q}}_s(N)
\underset{s\to\infty}{\sim}
\left[C_A\Gamma(A+1)\right]^N
\sum_{n=0}^N\binom{N}{n}\left(\frac{C_{A+1}(A+1)}{C_A}\right)^n\frac{1}{s^{N(A+1)+n+1}}
    \label{qsnlargests}
\end{equation}
The inverse Laplace of ${\hat{Q}}_s(N)$ yields
\begin{equation}
Q_t(N)\sim
\frac{\left[C_A\Gamma(A+1)\right) t^{A+1}]^N}{\Gamma\left(N(A+1)+1\right)}
\sum_{n=0}^N\frac{1}{n!}\left(\frac{C_{A+1}}{C_A}t\right)^n \left(\frac{\left(N(A+1)\right)! N!(A+1)^n}{(N-n)!\left(N(A+1)+n\right)!}\right).
    \label{qtnsummform}
\end{equation}
While Eq.~\eqref{qsnlargests} was developed in the limit of large $s$, we will show below that Eq.~\eqref{qtnsummform} holds for any fixed  $t$ in the limit when $N\to\infty$. By inspection it is easy to notice that the term $\left(N(A+1)\right)! N!(A+1)^n\big/(N-n)!\left(N(A+1)+n\right)!$ is very close to $1$ while $N$ is large and $n<<N$. Then in the limit when $N\to\infty$ we use the Taylor expansion $\exp(tC_{A+1}/C_a)=\sum_{n=0}^\infty (tC_{A+1}/C_a)^n/n!$ and write the asymptotic form of $Q_t(N)$
\begin{equation}
Q_t(N)\underset{N\to\infty}{\sim}
\displaystyle
\frac{\left(C_A\Gamma(A+1)t^{A+1}\right)^N}{\Gamma(N(A+1)+1)}
e^{\frac{C_{A+1}}{C_A}t}.
    \label{qtnassymptotic}
\end{equation}
A proper proof of Eq.~\eqref{qtnassymptotic} is provided in Sec.~\ref{qtnproofSec} . In Sec.~\ref{sechighorderpsi} we also develop the expression for $Q_t(N)$ when the next leading terms in $s$ of Eq.~\eqref{qsnexp02} are taken into account. 

Equation~\eqref{qtnassymptotic} describes the probability of obtaining large number of renewals ($N$) when the measurement time ($t$) is kept fixed. As we speculated, $Q_t(N)$ in this limit solely depends on the short $\tau$ properties of $\psi(\tau)$ while large $\tau$s are completely ignored. 
While the opposite limit of $t\to\infty$ is separated into three different classes (see Table~\ref{table:1}), for the $N\to\infty$ limit all these three classes can be united, as long as $\psi(\tau)$ is analytic in the vicinity of $0$, i.e. Eq.~\eqref{taylorexpansion}.
In Section~\ref{seclargedev} we use the language of the theory of Large Deviations in order to properly address these two limits and their universal features, as they emerge from Eq.~\eqref{qtnassymptotic} and Table~\ref{table:1}. 

\begin{figure}
   \centering
    \begin{subfigure}[b]{0.45\textwidth}
        \includegraphics[width=\textwidth]{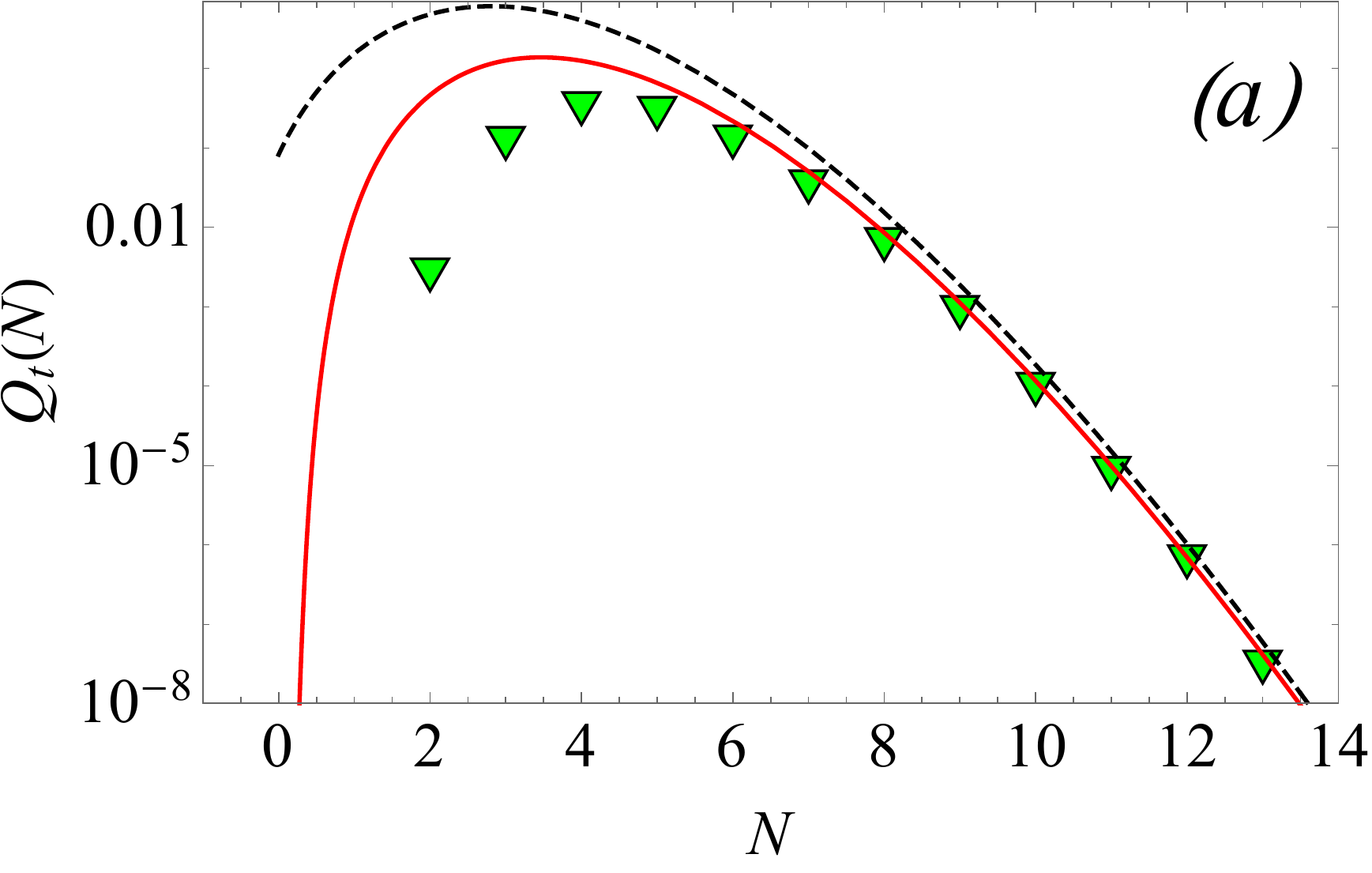}
    \end{subfigure}
    ~ 
    \begin{subfigure}[b]{0.45\textwidth}
        \includegraphics[width=\textwidth]{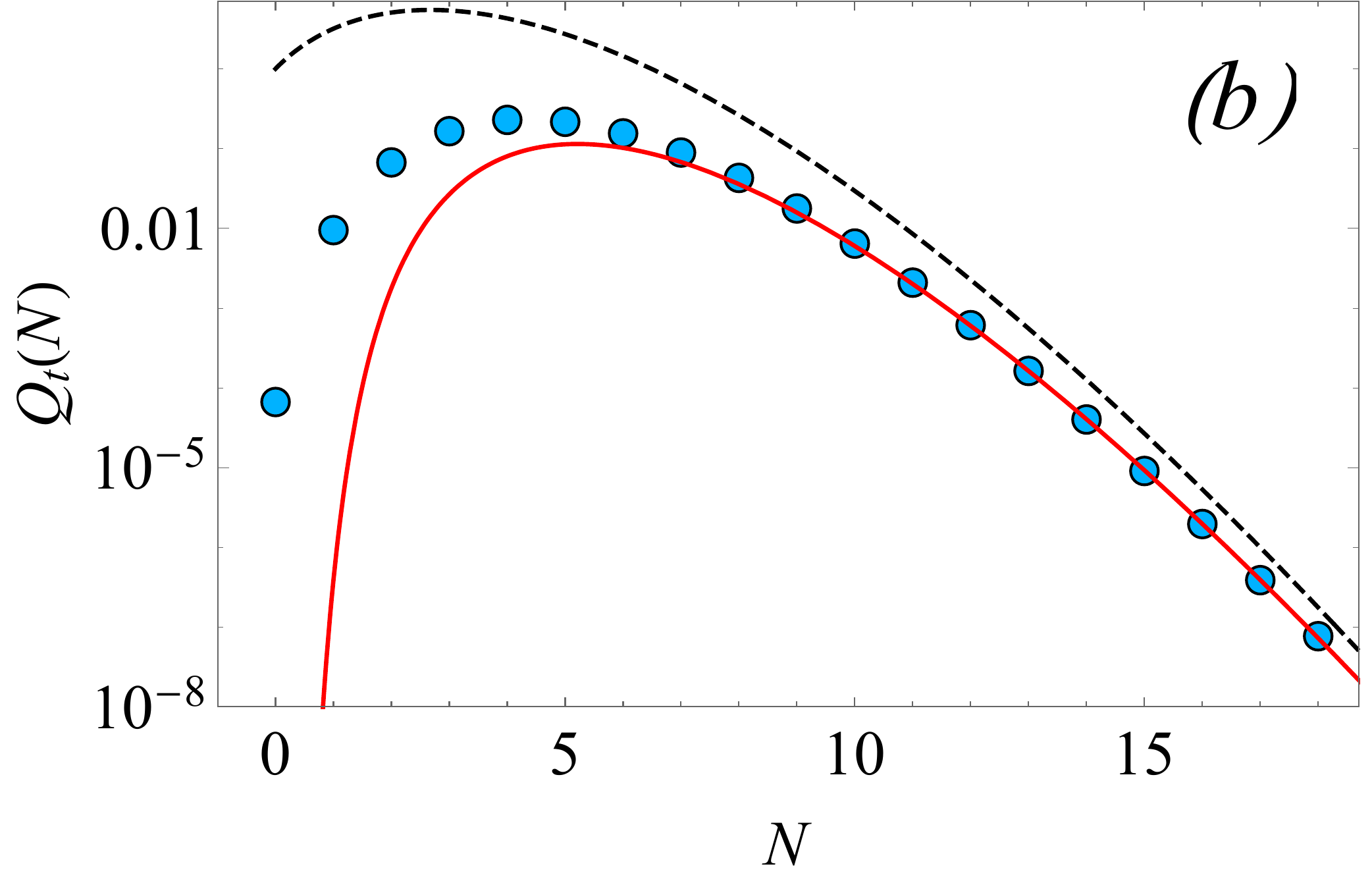}
    \end{subfigure}
    ~ 
    \begin{subfigure}[b]{0.45\textwidth}
        \includegraphics[width=\textwidth]{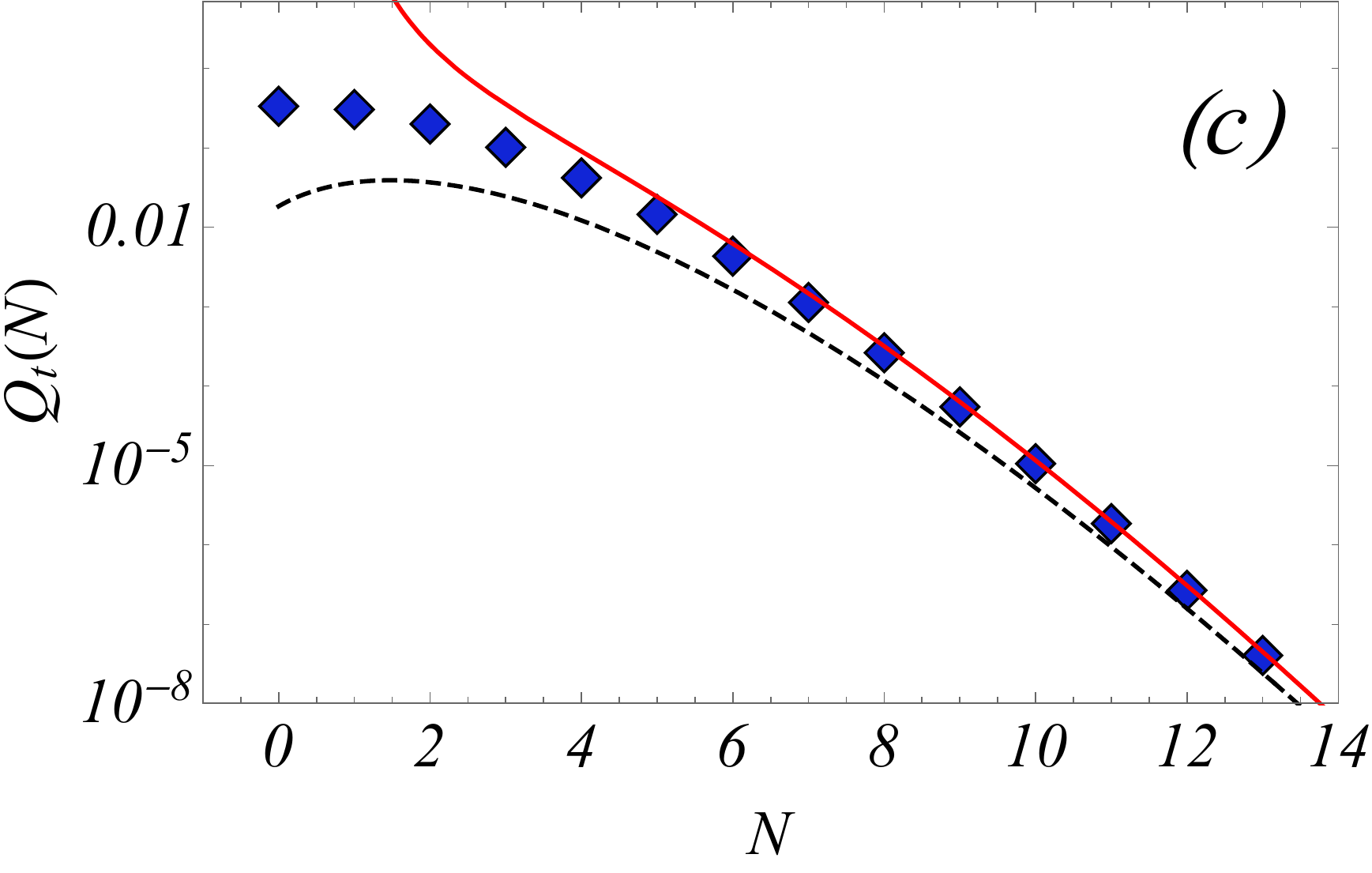}
    \end{subfigure}
    \caption{Comparison between the numerically obtained $Q_t(N)$ (symbols) and the analytic predictions of  Eq.~\eqref{qtnassymptotic} (dashed line) that describes the asymptotic behavior when $N\to\infty$ and Eq.~\eqref{qtncorrectiontn} (thick line) that include also corrections for finite $N$. Three different cases of $\psi(\tau)$ are considered: {\bf (a)} Beta distribution $\psi(\tau)=6\tau(1-\tau)$ with measurement time $t=2.5$, {\bf (b)} Half Normal distribution $\psi(\tau)=\frac{2}{\pi}e^{-\frac{x^2}{\pi}}$ with $t=5$ and {\bf (c)} Dagum distribution $\psi(\tau)=1/(1+\tau)^2$ with $t=2$.
    }
    \label{qtnexamples}
\end{figure}

For the case of exponential $\psi(\tau)$, Eq.~\eqref{qtnassymptotic} provides the same result as the exact form of $Q_t(N)$ in this case, i.e. Poisson distribution. Broadly speaking, the mathematical form in Eq.~\eqref{qtnassymptotic} is some sort of generalization of the Poisson form. The comparisons between the developed form of $Q_t(N)$ and several specific cases are provided in Fig.~\ref{qtnexamples}, together with the more precise forms of $Q_t(N)$ for finite $N$ that are developed in the next subsection. In the next sub-section we provide an explicit proof of Eq.~\eqref{qtnassymptotic} while the corrections and necessary condition for convergence are developed in Sec.~\ref{sechighorderpsi} (see Eq.~\eqref{qtncorrectiontn} and Eq.~\ref{qtnconvcondition}). 

\subsection{ Eq.~\eqref{qtnassymptotic} as a leading order in the expansion of $Q_t(N)$}
\label{qtnproofSec}

For the derivation of Eq.~\eqref{qtnassymptotic}  a hand-waving argument was utilized. While sufficient for initial introduction of the result, it is crucial to provide a rigorous derivation that also will be important for the development of corrections (for finite $N$) to the asymptotic result.

The leading order (in $N$) of $Q_t(N)$ in the $N\to\infty$ limit is provided by Eq.~\eqref{qtnsummform} and can also be written as 
\begin{equation}
{{Q}}_t(N)\underset{N\to\infty}{\sim}
\displaystyle
\frac{\left(C_A\Gamma(A+1)t^{A+1}\right)^N}{\Gamma(N(A+1)+1)}
\sum_{n=0}^N
\frac{\left[t\frac{C_{A+1}\Gamma(A+2)}{C_A\Gamma(A+1)}\right]^{n}}{n!}
\frac{\left(N(A+1)\right)! N!}{(N-n)!\left(N(A+1)+n\right)!}.
    \label{qtnlong03}
\end{equation}
For the fraction of factorials we use $N!/(N-n)!=(-1)^n(-(N-n+1))(-(N-n+2))\dots (-N)$ and 
$(N(A+1)+n)!/(N(A+1))!=(N(A+1)+1)\dots(N(A+1)+n)$ to obtain
\begin{equation}
    {{Q}}_t(N)\underset{N\to\infty}{\longrightarrow}
\displaystyle
\frac{\left(C_A\Gamma(A+1)t^{A+1}\right)^N}{\Gamma(N(A+1)+1)}
\sum_{n=0}^N
\frac{\left[-t\frac{C_{A+1}(A+1)}{C_A}\right]^{n}}{n!}\frac{(-N)_n}{(N(A+1)+1)_n},
    \label{qtnshort01}
\end{equation}
where $(a)_b=\Gamma(a+b)/\Gamma(a)=a(a+1)\dots(a+b-1)$ is the Pochhammer symbol~\cite{Abramowitz}. When $n>N$ the $(-N)_n$ will include a multiplier $(-N+N)=0$ and so for any $n>N$  $(-N)_n=0$ and we can extend the upper limit of summation in Eq.~\eqref{qtnshort01} up to $n\to\infty$. Then by using the definition of ${}_1F_1(a;b;z)$, i.e the Kummer function of the first kind~\cite{Abramowitz}, ${}_1F_1(a;b;z)=\sum_{n=0}^\infty(a)_n z^n\Big/(b)_n n!$, we obtain that 
\begin{equation}
    {{Q}}_t(N)\underset{N\to\infty}{\sim}
\displaystyle
\frac{\left(C_A\Gamma(A+1)t^{A+1}\right)^N}{\Gamma(N(A+1)+1)}
{}_1F_1\left(-N;N(A+1)+1;-t\frac{C_{A+1}(A+1)}{C_A}\right).
    \label{qtnshort02}
\end{equation}
This result is the leading $N$-order behavior of $Q_t(N)$ when $N\to\infty$. The proof that in the limit when $N\to\infty$ Eq.~\eqref{qtnshort02} converges to Eq.~\eqref{qtnassymptotic}, as was previously suggested, follows directly from the properties of Kummer function.
The Kummer function ${}_1F_1(a;b;z)$ satisfies the second order differential equation
\begin{equation}
z\frac{d^2{}_1F_1(a;b;z)}{dz^2}+(b-z)\frac{d{}_1F_1(a;b;z)}{dz}-a{}_1F_1(a;b;z)=0, 
\label{kummerdiff}
\end{equation}
that for the specific parameters $a=-N$ and $b=N(A+1)+1$ is given by 
\begin{equation}
    z\frac{d^2{}_1F_1}{dz^2}+(N(A+1)+1-z)\frac{d {}_1F_1}{dz}+N{}_1F_1=0.
    \label{kummerf}
\end{equation}
Multiplying Eq.~(\ref{kummerf}) by $1/N$ and taking the $N\to\infty$ limit we obtain $\frac{d {}_1F_1}{dz}+\frac{1}{A+1}{}_1F_1=0$ which provides the asymptotic behavior 
\begin{equation}
    {}_1F_1(-N;N(A+1)+1;z)\underset{N\to\infty}{\sim}
    \exp\left(-\frac{z}{A+1}\right).
    \label{kummersol}
\end{equation}
Eq.~(\ref{kummersol}) and Eq.~(\ref{qtnshort02}) give the asymptotic behavior of $Q_t(N)$ as  is provided by Eq.~\eqref{qtnassymptotic}.
~

In Fig.~\ref{qtnexamples} the comparison between Eq.~\eqref{qtnassymptotic} and numerical simulations is performed. We do see that the $Q_t(N)$ is indeed approaches to the theoretical result of Eq.~\eqref{qtnassymptotic} but some deviations are still noticeable. Specifically in panel {\bf (b)} where the expected number of renewals $t/\langle \tau \rangle$ is $5$ we see a difference even when $N>15$. This suggests that corrections for finite $N$ of the asymptotic form of Eq.~\eqref{qtnassymptotic} are needed, together with some sort of criteria of convergence to this asymptotic result. These issues are addressed in the next subsection.

\subsection{The effect of higher order expansion of $\psi(\tau)$}
\label{sechighorderpsi}

The full expansion of ${\hat{Q}}_s(N)$ in the limit of $s\to\infty$ is provided by Eq.~\eqref{qsnexp02}. Previously we explicitly used only the first two terms, i.e. $C_{A}$ and $C_{A+1}$, in the expansion of ${\hat{\psi}}(s)$ and it is not clear how the next terms in the expansion affect the result. From the form of Eq.~\eqref{qsnexp02} it is hard to guess how the corrections are going to behave. Moreover,  a clear criteria as for where to cut the expansion of ${\hat{\psi}}(s)$ is needed. To accomplish this task we rewrite  Eq.~\eqref{qsnexp02} as
\begin{equation}
    {\hat{Q}}_s(N)=\frac{\left[C_A\Gamma(A+1)\right]^N}{s^{N(A+1)+1}}\sum_{n=0}^N\binom{N}{n}\left(\sum_{i=1}^\infty\frac{C_{A+i}\Gamma(A+i+1)}{C_A \Gamma(A+1) s^{i}}\right)^n
\left(1-\sum_{i=0}^\infty\frac{C_{A+i}\Gamma(A+i+1)}{s^{A+i+1}}\right)
    \label{qsnexp03}
\end{equation}
while using the binomial expansion. Next, by applying the multinom expansion $(\sum_{i=0}^m w_i )^n= \sum_{\{k_i\}}\frac{n!}{{ \prod_{i=0}^m k_i!}}\prod_{i=0}^m w_i^{k_i}$,  the summation is for every possible set of values for $k_i$ (integer) that satisfy the condition $\displaystyle\sum_{i=0}^m k_i=n$. In the $m\to\infty$ limit  Eq.~\eqref{qsnexp03} yields
\begin{equation}
\begin{array}{l}
{\hat{Q}}_s(N)=\displaystyle \sum_{n=0}^N\sum_{\{ k_i\}}\sum_{j=0}^\infty
\binom{N}{n} 
\frac{\left[C_A\Gamma(A+1)\right]^N}{s^{N(A+1)+1}}\times
\\
\times \displaystyle
\frac{n!}{{\displaystyle \prod_{i=0}^\infty k_i!}}\left(\prod_{i=0}^\infty \left[\frac{C_{A+1+i}\Gamma(A+2+i)}{C_A\Gamma(A+1)}\right]^{k_i}\right)\frac{B_j}{s^{\sum_{i=0}^\infty(i+1)k_i+l_j}},
\end{array}
    \label{qsnexp04}
\end{equation}
where $\forall j>0$, $B_j=-C_{A+j-1}\Gamma(A+j)$ , $l_j=A+j$ and $B_0=1$ , $l_0=0$. By taking inverse Laplace transform term by term in Eq.~\eqref{qsnexp04} we finally obtain
\begin{equation}
\begin{array}{l}
    {{Q}}_t(N)=
\displaystyle
\frac{\left(C_A\Gamma(A+1)t^{A+1}\right)^N}{\Gamma(N(A+1)+1)}
\sum_{n=0}^N\sum_{\{ k_i\}}\sum_{j=0}^\infty
B_j
t^{ \displaystyle \left[  \sum_{i=0}^\infty(i+1)k_i+l_j \right] }\times
\\
\times
\displaystyle
\left(\prod_{i=0}^\infty \left[\frac{C_{A+1+i}\Gamma(A+2+i)}{C_A\Gamma(A+1)}\right]^{k_i}\right)
\frac{\left(N(A+1)\right)! N!}{(N-n)!{\displaystyle \prod_{i=0}^\infty k_i!}}\times
\frac{1}{\displaystyle \Gamma\left(N(A+1)+\sum_{i=0}^\infty(i+1)k_i+l_j+1\right)},
\end{array}
    \label{qtnlong02}
\end{equation}
where again the summation $\sum_{\{k_i\}}$ is for every possible set of values for $k_i$ (integer) that satisfy the condition $\sum_{i=0}^\infty k_i=n$. 
Next we change the order of summation in Eq.~\eqref{qtnlong02}. Instead of first summing up  all given realizations $\{k_i\}$ that satisfy $\sum_i^\infty k_i=n$, we first want to sum  all different $n$s for a specific realization $\{k_i\}_{\kappa1,\kappa2}$. 
The realization $\{k_i\}_{\kappa1,\kappa2}$ is defined as a realization for which the lowest possible index $i$, for which $k_i>0$, is greater than $0$ while $\sum_{i=1}^\infty k_i=\kappa1$ and  $\sum_{i=m+1}^\infty i k_i=\kappa2$. We notice that 
\begin{equation}
\begin{array}{l}
\displaystyle
{
{}_1 F_1\left(-(N-\kappa1);N(A+1)+\kappa1+\kappa2+l_j+1;-\frac{C_{A+1}}{C_A}(A+1)t\right)
}
\\
\displaystyle
=\left(\frac{N!N((A+1))!}{(N-\kappa1)!(N(A+1)+\kappa1+\kappa2+l_j)!}\right)^{-1}
\sum_{n=\kappa1}^N t^{(n-\kappa1)}
\left(\frac{C_{A+1}\Gamma(A+2)}{C_A\Gamma(A+1)}\right)^{n-\kappa1}
\\
\displaystyle
\times
\frac{\left(N(A+1)\right)! N!}{(N-n)! (n-\kappa1)!\Gamma\left(N(A+1)+\kappa1+\kappa2+(n-\kappa1)+l_j+1\right)},
\end{array}
    \label{hypgemsum01}
\end{equation}
and Eq.~\eqref{qtnlong02} can be rewritten as  
\begin{equation}
\begin{array}{l}
 {{Q}}_t(N)=
\displaystyle
\frac{\left(C_A\Gamma(A+1)t^{A+1}\right)^N}{\Gamma(N(A+1)+1)}
\sum_{j=0}^\infty
\sum_{\kappa1=0}^N\sum_{\kappa2}\sum_{\{k_i\}_{\kappa1,\kappa2}}
 \frac{ B_jt^{\kappa1+\kappa2+l_j}N!N((A+1))!b(\{k_i\}_{\kappa1,\kappa2})}
 {(N-\kappa1)!(N(A+1)+\kappa1+\kappa2+l_j)!}
\\
\times
\displaystyle
{
{}_1 F_1\left(-(N-\kappa1);N(A+1)+\kappa1+\kappa2+l_j+1;-\frac{C_{A+1}}{C_A}(A+1)t\right)
},
\end{array}
    \label{qtngenkummer01}
\end{equation}
where $b(\{k_i\}_{\kappa1,\kappa2})=\left(\prod_{i=1}^\infty\frac{1}{k_i!} \left[\frac{C_{A+1+i}\Gamma(A+2+i)}{C_A\Gamma(A+1)}\right]^{k_i}\right)$ and $\forall j>0$, $B_j=-C_{A+j-1}\Gamma(A+j)$ , $l_j=A+j$ and $B_0=1$ , $l_0=0$. 
The summation $\sum_{k2}$ is constrained to all the values $\sum_{i=1}^\infty ik_i=\kappa2$ while  $\sum_{i=1}^\infty k_i=\kappa1$ . The summation $\sum_{\{k_i\}_{\kappa1,\kappa2}}$ is for all possible realizations of $k_i$ for given $\kappa1$ and $\kappa2$. The advantage of writing $Q_t(N)$ as a sum of Kummer functions is the fact that the asymptotic expression for ${}_1 F_{1}$ (when the parameters are as in Eq.~\eqref{qtngenkummer01}) is obtained in a quite straightforward manner (see  Appendix~\ref{appendix:hqfp}). The pre-factor of ${}_1F_{1}$ includes an integer power of $t$ and a quotient of factorials that can be expanded in a power-series of $1/N$.
Indeed, the powers of $t/N^\beta$ in the pre-factor of ${}_1F_1$ appear due to the terms $t^{\kappa1+\kappa2+l_j}N!(N(A+1))!/(N-\kappa1)!(N(A+1)+\kappa1+\kappa2+l_j)!$. 
For sufficiently large $N$ this term transforms into $t^{\kappa1+\kappa2+l_j}/N^{\kappa2+l_j}$. Since $\kappa2>\kappa1$ when $k_i>0$ for any $i>1$, we notice that only such combinations  $\{k_i\}_{\kappa1,\kappa2}$ where $\forall i>1$ $k_i=0$ contribute to terms  with integer powers of $(t/N^{1/2})$. This is due to the fact that for this case $\kappa1=\kappa2$.  Any other combination of $\{k_i\}_{\kappa1,\kappa2}$ produce integer powers of $(t/N^\beta)$ with $\beta>1/2$. 
According to Appendix~\ref{appendix:hqfp} the function ${}_1 F_1$ is also a sum of terms that are multiplied by integer powers of $t/N^\beta$. Summing up all the terms up to $\beta\leq1/2$ produces (see Eq.~\eqref{appkummernpower}) the approximation
\begin{equation}
\displaystyle
{
{}_1 F_1\left(-(N-\kappa1);N(A+1)+\kappa1+\kappa2+l_j+1;-\frac{C_{A+1}}{C_A}(A+1)t\right)
}
\underset{N\to\infty}{\sim}
\displaystyle{e^{\frac{C_{A+1}}{C_A}t-\frac{(A+2)C_{A+1}^2}{2(A+1)C_A^2}\frac{t^2}{N}}}
    \label{1f1approximation}
\end{equation}
We notice that
\begin{equation}
\begin{array}{l}
\displaystyle
{}_1 F_2\left(-N;\frac{N(A+1)+1}{2},\frac{N(A+1)+2}{2};-\frac{C_{A+2}(A+1)(A+2)}{4C_A}t^2 \right)=
\\
\displaystyle
\sum_{k=0}^N 
\frac{ t^{2k}N!N((A+1))!}
 {(N-\kappa1)!(N(A+1)+\kappa1+\kappa2+l_j)!k!}
 \left(\frac{C_{A+2}(A+1)(A+2)}{C_A}\right)^k,
  \end{array}
    \label{1f2approximation}
\end{equation}
where ${}_1 F_2$ is the generalized hypergeometric function (see Appendix~\ref{appendix:hg1f2}), and all the pre-factors of ${}_1 F_1$ in Eq.~\eqref{qtngenkummer01} that include integer powers of $t/N^{1/2}$ are summed up and produce this ${}_1 F_2$ function in Eq.~\eqref{1f2approximation}. We use Eq.~\eqref{1f1approximation}, Eq.~\eqref{1f2approximation} and its large $N$ approximation (see Eq.~\eqref{apphg1F2Nlarge}) and obtain from Eq.~\eqref{qtngenkummer01} the expression for $Q_t(N)$ that includes the contribution of all terms of the type $t/N^\beta$, $\forall \beta \leq 1/2$
\begin{equation}
Q_t(N)\underset{N\to\infty}{\sim}
\frac{\left(C_A\Gamma(A+1)t^{A+1}\right)^N}{\Gamma(N(A+1)+1)}
\displaystyle{
e^{\frac{C_{A+1}}{C_A}t+\frac{(A+2)}{(A+1)}
\left( \frac{C_{A+2}}{C_A}-\frac{1}{2}\frac{C_{A+1}^2}{C_A ^2}
\right)\frac{t^2}{N}}
}.
    \label{qtncorrectiontn}
\end{equation}
Eq.~\eqref{qtncorrectiontn} includes the corrections to the asymptotic form of $Q_t(N)$, as it is displayed by Eq.~\eqref{qtnassymptotic}. From the derivation of Eq.~\eqref{qtncorrectiontn} it become clear that any additional correction to the asymptotic for of $Q_t(N)$ will involve terms with $t/N^\beta$ with $\beta>1/2$ that will decay to $0$ when $N\to\infty$ faster than the terms that appear in Eq.~\eqref{qtncorrectiontn}. We are now ready to state the necessary condition for convergence of $Q_t(N)$ to the asymptotic form in Eq.~\eqref{qtnassymptotic}, i.e.
\begin{equation}
\frac{t}{N} \ll \Big| \frac{2(A+1)C_A C_{A+1}}{(A+2)[2C_{A+2}C_A-C_{A+1}^2]}\Big|.
    \label{qtnconvcondition}
\end{equation}
This condition is true when the coefficient $C_{A+1}\neq 0$.
Otherwise $[(A+1)C_{A+2}\Big/(A+1)C_A]t^2/N$ should be small.

During the derivation of Eq.~\eqref{qtncorrectiontn} all terms that included powers of $t/N^\beta$, with $\beta>1/2$, were neglected. The properties of the the hyper-geometric functions, briefly introduced in the Appendixes, suggest that all those additional terms  will introduce multiplicative factors in Eq.~\eqref{qtncorrectiontn}. These multiplicative pre-factors will  converge to $1$ in the large $N$ limit. 
In other words, the additional corrections are expected to contribute additional terms in the exponent function in Eq.~\eqref{qtncorrectiontn}, those terms are expected to decay faster than $t^2/N$ since they are functions of $t/N^\beta$ with $\beta>1/2$.

In Figure~\ref{qtnexamples} we see that the prediction that include corrections for finite $N$, i.e. Eq.~\eqref{qtncorrectiontn}, is superior to the asymptotic prediction for the explored cases. For large enough $N$ both expression converge to exactly the same result.  
For sufficiently low $N$ we do see noticeable differences between the two expressions and the faster convergence of the result with higher correction terms, i.e. Eq.~\eqref{qtncorrectiontn}, to the true behavior. It is interesting to notice that for the case when $\psi(\tau)$ is exponential, i.e. $\psi(\tau)=\exp(-\tau/\langle \tau \rangle)/\langle \tau \rangle$, Eq.~\eqref{qtnassymptotic} is the exact $Q_t(N)$ and indeed the correction term in Eq.~\eqref{qtncorrectiontn} simply zeros out.

While the comparison of the derived results for $Q_t(N)$ with real examples was performed for not large measurement time $t$, the formulas are not limited to small times. The only limitation is smallness of $t/N$, as it is expressed  by Eq.~\eqref{qtnconvcondition}. This leads us to conclusion that the behavior of $Q_t(N)$ in the large $N$ limit can be connected to the behavior of this function in the $t\to\infty$ limit, i.e.  the formulas in Table~\ref{table:1}, especially the \# I case. We address this task in the next section by utilizing the language of Large Deviations.

\begin{figure} 
\centering
			\includegraphics[width=0.7\textwidth]{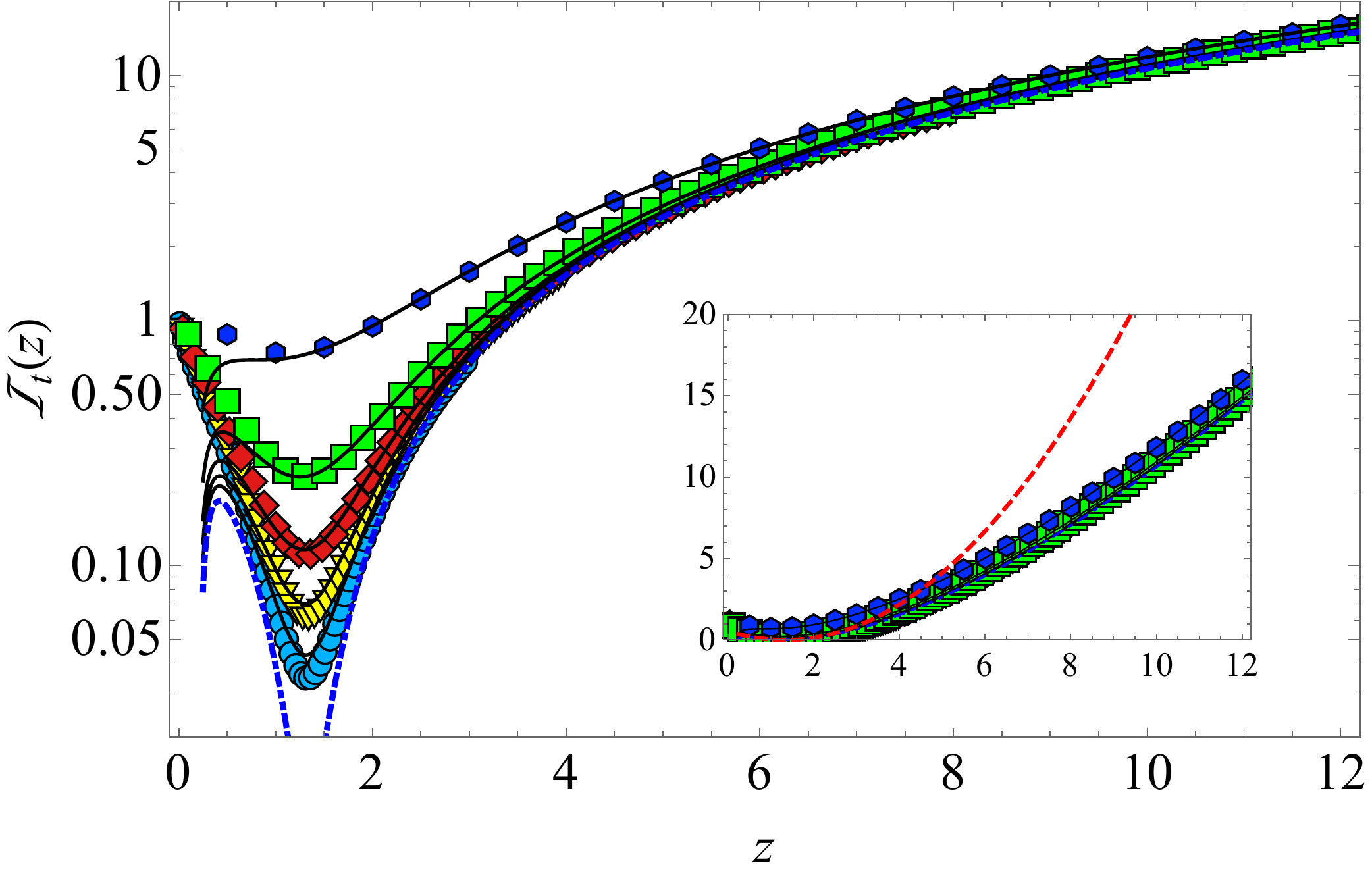}
\caption{
The rate function ${\cal I}_t(z)$ for the case of $\psi(\tau)=\frac{1}{2}\left(e^{-\tau}+2e^{-2\tau}\right)$ and several measurement times $t$. 
The symbols are numerical realizations of ${\cal I}_t(z)$  for: $t=2$ ($\varhexagon$), $t=10$ ($\Square$), $t=25$ ($\diamond$), $t=50$ ($\triangledown$) and $t=100$ ($\Circle$). 
Thick lines are the theoretical predictions that use Eq.~\eqref{qtncorrectiontn} for $Q_t(z t)$ in the ${\cal I}_t(z)$ definition (Eq.~\eqref{ratefinitt}). 
The dashed-dotted line describe the asymtotic rate function ${\cal I}_t(z)$ , as it is provided by Eq.~\eqref{ratefunclrgNz}. In the inset the same figure is presented bu with linear scale for the $y$ axis. The addition is the dashed line that describes the Gaussian behavior of ${\cal I}(z)$, according to Eq.~\eqref{ratefuncgaussfin}. 
}
    \label{rateconexample}
\end{figure}

\section{Large Deviations and the Rate Functions}
\label{seclargedev}

The behavior of $Q_t(N)$ in the limit of large measurement time $t$ can be described by using the language of Large Deviations~\cite{Touchete,Derrida,Dhar,Majumdar,TouchetteA} that has been applied to renewal-reward process (i.e CTRW) ~\cite{ChiLarge,Lefevere,Tsirelson} and general Markov processes~\cite{Lapolla2018Unfolding}. The Large Deviations Principle states that in the limit of $t\to\infty$ when taking a fixed ratio $N/t=z$, $Q_t(N)$ can be described by the means of a rate function ${\cal I}(z)$, i.e. $Q_t(N)\sim e^{-t{\cal I}(z)}$. For any finite $t$  the rate function is defined as
\begin{equation}
{\cal I}_t(z)=-\log\left( Q_t(N) \right)\Big/t
    \label{ratefinitt}
\end{equation}
when we set $z=N/t$.
One can find the rate function when the distribution of the random variables ($\tau$s) is not broad, i.e. case \# I in Table~\ref{table:1}.
Specifically one can deduce that due to the Gaussian property of $Q_t(N)$ there is a universal behavior of ${\cal I}(z)$.
Indeed, taking the $t\to\infty$ limit in Eq.~\eqref{ratefinitt} and using the Gaussian form of $Q_t(N)$ in Eq.~\eqref{qtngaussianfinal} we obtain that
\begin{equation}
{\cal I}(z) \sim \frac{(z-1/\mu)^2}{2\sigma^2/\mu^3} 
    \label{ratefuncgaussfin}
\end{equation}
when $z=N/t$ is in the vicinity of $1/\mu$.
The rate function ${\cal I}(z)$ states that the PDF of the random variable $z=N_t/t$ is universal in the $t\to\infty$ limit in the vicinity of $z=1/\mu$ when there exist a finite mean and variance for the $\psi(\tau)$. 
While this fact is commonly known, the presented results for the behavior of $Q_t(N)$ in the $N\to\infty$ limit suggest that there is an additional universal behavior of ${\cal I}(z)$ irrespective of the existence of average $\tau$ (i.e. finite $\mu$). 
The limit of large $N$ of $Q_t(N)$, as it is expressed by Eq.~\eqref{qtncorrectiontn}, is not assuming small $t$. Thus 
by using Stirling's approximation in the $N\to\infty$ limit and keeping $N$ such that the condition in Eq.~\eqref{qtnconvcondition} is satisfied , from Eq.~\eqref{qtncorrectiontn} we obtain that
\begin{equation}
\begin{array}{l}
\displaystyle
{\cal I}_t(z)\sim
\\
\displaystyle
(A+1)\frac{N}{t}\log\left(\frac{N}{t}\right)+
\frac{N}{t}\log\left(\frac{(A+1)^{A+1}}{C_A\Gamma(A+1)e^{A+1}}\right)
-\frac{C_{A+1}}{C_A}
-\frac{(A+2)}{(A+1)}
\left( \frac{C_{A+2}}{C_A}-\frac{1}{2}\frac{C_{A+1}^2}{C_A ^2}
\right)\frac{1}{\frac{N}{t}}
\end{array}
    \label{ratefunclrgN01}
\end{equation}
or in terms of  the rate function
\begin{equation}
{\cal I}(z)\sim
z\left[(A+1)\log\left(z\right)+
\log\left(\frac{(A+1)^{A+1}}{C_A\Gamma(A+1)e^{A+1}}\right)\right]
-\frac{C_{A+1}}{C_A}
-\frac{(A+2)}{(A+1)}
\left( \frac{C_{A+2}}{C_A}-\frac{1}{2}\frac{C_{A+1}^2}{C_A ^2}
\right)\frac{1}{z}
    \label{ratefunclrgNz}
\end{equation}
in the large-$z$ limit. We see that indeed in the functional form of ${\cal I}(z)$ in the large $z$ limit is universal in the same sense that the form of the rate function is universal near its minima. Meaning, a parabolic behavior for $z=1/\mu$ and almost linear (with logorithmic corrections) behavior for large enough $z$. While the condition for parabolic behavior is the existence of mean and variance for $\psi(\tau)$, the large $z$ universality need  $\psi(\tau)$ to be analytic in the vicinity of $\tau=0$. 
The transition between the large $z$ limit of ${\cal I}(z)$ and the parabolic behavior near $z\sim 1/\mu$ is not universal and depends on all the properties of $\psi(\tau)$.
Already in the form of ${\cal I}(z)$, as it appears in Eq.~\eqref{ratefunclrgNz}, there is a term that is proportional to $1/z$. This term is due to the corrections to $Q_t(N)$ that were introduced in Sec.~\ref{sechighorderpsi} by inclusion of powers of $t/N^{1/2}$. If additional correction terms for $Q_t(N)$ of the form $t/N^\beta$ ($\beta>1/2$) will be included it will lead to higher powers of $1/z$ in the form of ${\cal I}(z)$ and also appearance of higher orders of the Taylor expansion of $\psi(\tau)$. Those terms are negligible in the large $z$ limit while for small enough $z$ they contribute in a non-trivial way to the appearance of parabolic behavior around $z=1/\mu$.

\begin{figure} 
\centering
			\includegraphics[width=0.7\textwidth]{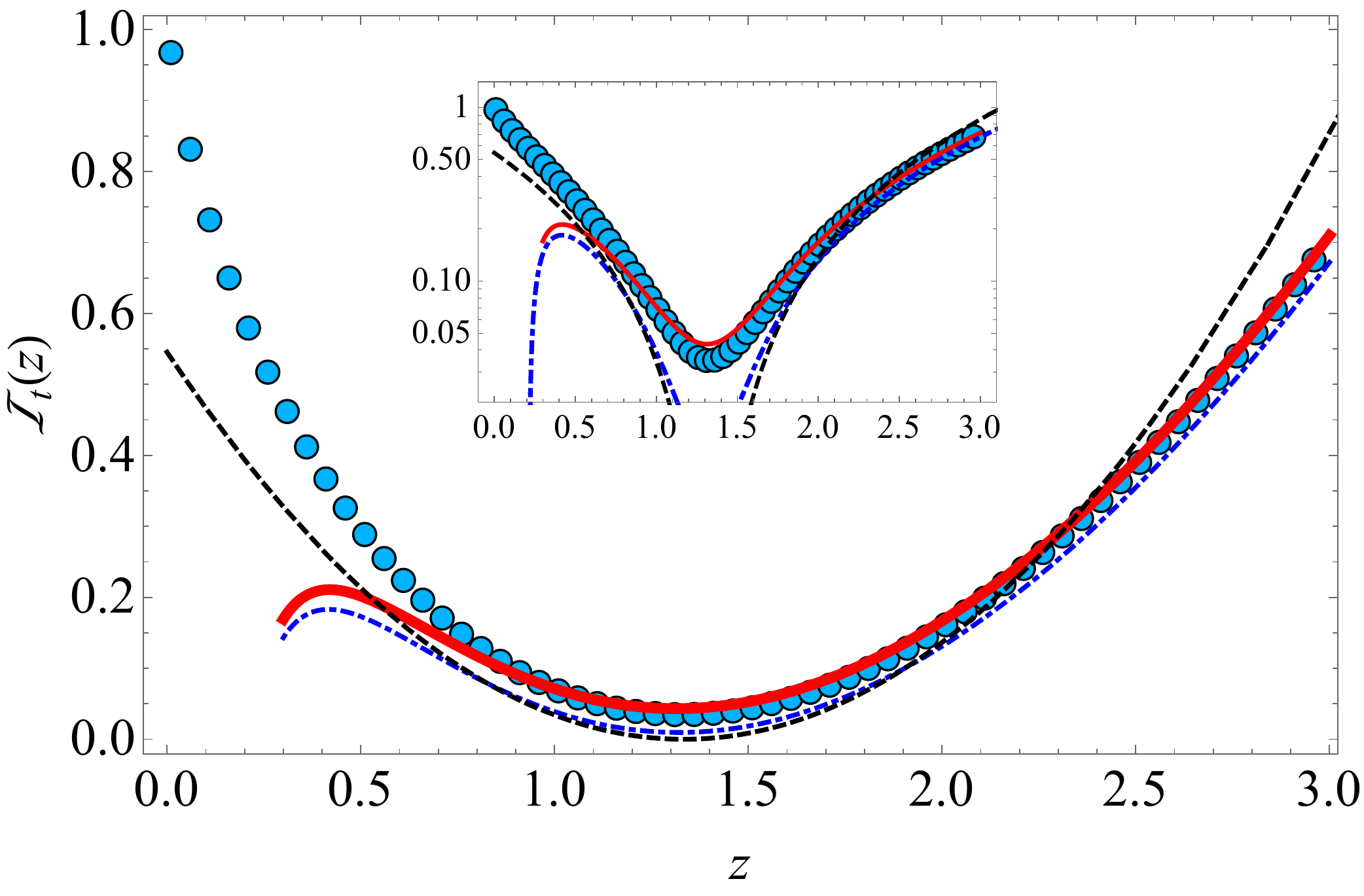}
\caption{
The rate function ${\cal I}_t(z)$ for the case of $\psi(\tau)=\frac{1}{2}\left(e^{-\tau}+2e^{-2\tau}\right)$ and $t=100$. Numerical realization is described by symbols. The thick line is the theoretical prediction for ${\cal I}_{100}(z)$ and the dashed-dotted line is the asymptotic behavior of ${\cal I}(z)$ according to Eq.~\eqref{ratefunclrgNz}. The dashed line displays the theoretical prediction according to the parabolic form of ${\cal I}(z)$, i.e. Eq.~\eqref{ratefuncgaussfin}. The inset displays the same plot with logarithmic scale of the $y$ axis and focuses on the discrepancies near $z=1/\mu$.
}
    \label{ratet100example}
\end{figure}

The convergence of the rate function to both universal forms is displayed in Fig.~\ref{rateconexample} and Fig.~\ref{ratet100example}  for the specific case of $\psi(\tau) = \frac{1}{2}e^{-\tau}+e^{-2\tau}$. The convergence to a parabola occurs for $z\sim 1/\mu=4/3$ and for $z>2$ we already start to see  the convergence to the universal limit for large $z$. 
From this comparison we can also see the crucial difference between the two universal limits. While parabolic behavior behavior occurs only in the large $t$ limit, the large $z$ form of ${\cal I}(z)$ is not limited to large $t$.
For Eq.~\eqref{qtncorrectiontn} to work properly the condition $t/N<<1$ (see Eq.~\eqref{qtnconvcondition}) must be satisfied, and $t$ must not be large. This is why the rate functions that are derived for finite time work extremely well for the examples in Fig.~\ref{rateconexample} and Fig.~\ref{ratet100example}. 
For the described example we see (Fig.~\ref{rateconexample}) that already at $t=10$ the convergence at the tails to asymptotic ${\cal I}(z)$ is extremely good (the ${\cal I}_t(z)$ works perfectly even when $t=2$). 
The discrepancies between the true behavior and the asymptotic ${\cal I}(z)$ around $z=4/3$ are still noticeable even for $t=100$ (see Fig.~\ref{ratet100example}). 

\section{Summary}
\label{secsummary}

The number of renewals is an important quantity in many physical and mathematical models where the renewal assumption holds. We focused in this work on the behavior of the distribution of the number of renewals ($N$), when $N$ is large and the measurement time $t$ is fixed. 
While the regular assumption is that universal behavior of $Q_t(N)$ (summarized in Table~\ref{table:1}) appears only for large $t$ we show that in the limit of large $N$ there is an additional universal limit that holds even when $t$ is small. 
The only condition of this limit is the analyticity of the distribution of the waiting times, $\psi(\tau)$ in the vicinity of $\tau=0$.
For large $t$ the universality is displayed as a general functional form of $Q_t(N)$, e.g. Gaussian or L\'{e}vy, that differs from one $\psi(\tau)$ to another only by the values of such components as average $\tau$, the variance or the power-law decay of $\psi(\tau)$. When $N$ is large the functional form is always the same (Eq.~\eqref{qtnassymptotic}) and the difference between different $\psi(\tau)$s is achieved via the difference in the coefficients of Taylor expansion of $\psi(\tau)$ that enter the formula.
By exploration of the properties of hyper-geometric functions the corrections for $Q_t(N)$, for any finite $N$, were provided (Eq.~\eqref{qtncorrectiontn}) and the necessary condition for convergence to the asymptotic form  was  obtained (Eq.~\eqref{qtnconvcondition}).
We also extended our results to the limit of large measurement times. Using the language of the theory of Large Deviations for the number of renewals, we show that the form of the Large Deviations rate function ${\cal I}(z)$ is always linear (up to logarithmic corrections) in the large $z$ limit, as opposed to the parabolic behavior that is present in the vicinity of $z=1/\mu$ (only for case \#I of Table~\ref{table:1}).

It is finally important to notice that while we explored the behavior of the random variable $N_t$, the equations that were utilized are similar to the equations that define the probability distribution of $t_N$. Specifically, the cumulative distribution of the time $t$ that will take $N$ renewals to occur is provided by Eq.~\eqref{qtnassymptotic} and Eq.~\eqref{qtncorrectiontn}, in the large $N$ limit.

{\bf Acknowledgments:} This work was supported by the Pazy Foundation grant 61139927. I thank E. Barkai for stimulating discussions.

\appendix

\section{The Asymptotic Behavior of the Kummer Function of the First Kind}
\label{appendix:hqfp}

 The Kummer function of the first kind~\cite{Abramowitz} is defined as
 \begin{equation}
 {}_1F_1(a;b;z)=\sum_{n=0}^\infty(a)_n z^n\Big/(b)_n n!
     \label{appkummerdef}
 \end{equation}
where 
\begin{equation}
(a)_b=\frac{\Gamma(a+b)}{\Gamma(a)}=a(a+1)\dots(a+b-1)
    \label{pochamerdef}
\end{equation}
 is the Pochhammer symbol~\cite{Abramowitz}. ${}_1F_1(a;b;z)$ satisfies the second order differential equation
\begin{equation}
z\frac{d^2{}_1F_1(a;b;z)}{dz^2}+(b-z)\frac{d{}_1F_1(a;b;z)}{dz}-a{}_1F_1(a;b;z)=0, 
\label{appkummerdif}
\end{equation}
that for a specific choice of the parameters $a=-(N-d_1)$ and $b=N(A+1)+d_2+1$ (for  integer $d_1$ and $d_2$) turns into 
\begin{equation}
\epsilon z \frac{d^2y}{dy^2}+\left(A+1+\epsilon(1+d_2-z)\right)\frac{dy}{dz}+(1-\epsilon d_1)y=0.
    \label{kummerepsilon}
\end{equation}
for $y={}_1 F_1(-N;N(A+1)+1;z)$ and $\epsilon = 1/N$.
Since we are interested in the the behavior of ${}_1 F_1(-(N+d_1);N(A+1)+d_2+1;z)$ in the $N\to\infty$ limit, we use perturbation series for small $\epsilon$. Specifically the solution is presented as $y(z)=\sum_{k=0}^\infty \epsilon^k y_k(z)$. We use the boundary conditions $y_0(0)=1$ and $\forall k\geq 1$ $y_k(0)=0$. Introduction of this representation of $y(z)$ into Eq.~\eqref{kummerepsilon}  yields the relation
\begin{equation}
(A+1)\frac{dy_0}{dz}+y_0=0
    \label{yzerodiff}
\end{equation}
and
\begin{equation}
(A+1)\frac{dy_k}{dz}+y_k+z\frac{d^2 y_{k-1}}{dz^2}+(d_2+1-z)\frac{dy_{k-1}}{dz}-d_1 y_{k-1}=0\qquad\left(k>0\right).
    \label{ynonzerodiff}
\end{equation}
The solution for $y_0$ is similar to the asymptotic behavior that we already obtained in Eq.~\eqref{kummersol}, i.e. $y_0(z)=\exp(-z/(A+1))$. For any $k>0$ the recurrence relation in Eq.~\ref{ynonzerodiff} produces
\begin{equation}
y_k(z)=-\frac{e^{-\frac{z}{A+1}}}{A+1}\left[
\int e^{\frac{z}{A+1}}\left( z \frac{d^2 y_{k-1}}{dz^2}+(d_2+1-z)\frac{dy_{k-1}}{dz} -d_1 y_{k-1}\right)d\,z+\tilde{C}_k
\right]
\qquad(k>0),
    \label{ynonzeroint}
\end{equation}
where $\tilde{C}_k$ is a constant that is determined by applying the initial condition $y_k(0)=0$. The form of $y_0$ dictates that $\forall k$ $\tilde{C}_k=0$. 
From Eq.~\eqref{ynonzeroint} we can obtain the corrections up to any order, specifically 
\begin{equation}
y_1(z) = \frac{1}{(A+1)^3}e^{-\frac{z}{A+1}}
\left(-\frac{1}{2}(A+2)z^2+(A+1)(d_1(A+1)+d_2+1)z\right),
    \label{yfirstsecond}
\end{equation}
and for a general $k>0$
\begin{equation}
y_k(z) = \frac{1}{(A+1)^{3k}}e^{-\frac{z}{A+1}}
p_{2k}(z)
    \label{ygenksol}
\end{equation}
where $p_{2k}(z)$ is a polynomial  of order $2k$ without a constant term. For $p_{2k}(z)$ the leading non-constant term is $(-1)^{k}(A+2)^k z^{2k}/(2k)!!$, where $(2k)!!$ is the double factorial $(2k)!!=\prod_{j=1}^k(2j)$. Finally we can write the expansion of ${}_1 F_1(-(N+d_1);N(A+1)+d_2+1;z)$ as
\begin{equation}
{}_1 F_1(-(N+d_1);N(A+1)+d_2+1;z)=e^{-\frac{z}{A+1}}\left[1+\sum_{k=1}^\infty \frac{1}{N^k} \frac{p_{2k}(z)}{(A+1)^{3k}} \right].
    \label{appkummerexp}
\end{equation}

It is important to notice that since the leading order of $p_{2k}(z)$ is $2k$ then the highest ratio of the power of $z$ and the power of $N$, in the expansion of the Kummer function, is always $2$. This fact will become important in the main text when addressing the convergence of $Q_t(N)$ to its limit form.
Specifically, when taking into account only the terms that are powers of $t^2/N$ in Eq.~\eqref{appkummerexp} and using the fact that $\exp(-z^2/2)=\sum_{k=0}^\infty (-1)^k z^{2k}/(2k)!!$ we obtain that
\begin{equation}
\displaystyle
{}_1 F_1(-(N+d_1);N(A+1)+d_2+1;z)\underset{N\to\infty}{\sim}
\displaystyle{
e^{-\frac{1}{A+1}z-\frac{A+2}{2(A+1)^3}\frac{z^2}{N}}
}
    \label{appkummernpower}
\end{equation}

\section{The Asymptotic Behavior of the Generalized Hyper Geometric Function ${}_1 F_2$}
\label{appendix:hg1f2}

The generalized hypergeometric function ${}_1 F_2(a;b_1,b_2;z)$ is defined as
\begin{equation}
{{}_1 F_2(a;b_1,b_2;z)}=\sum_{n=0} \frac{1}{n!}\frac{(a)_n}{(b_1)_n (b_2)_n} \frac{z^n}{n!}
    \label{apphg1f2def}
\end{equation}
and it satisfies the third-order ordinary differential equation
\begin{equation}
z^2\frac{d^3 }{dz^3}{{}_1 F_2} +
(1+b_1+b_2)z\frac{d^2}{dz^2} {{}_1 F_2} +
\left[(b_1-1)(b_2-1)+b_1+b_2-1-z \right]\frac{d}{dz} {{}_1 F_2}
-a{{}_1 F_2}=0.
    \label{apphg1f2diff}
\end{equation}
Of specific interest are the parameters $a=-N$, $b_1=(N(A+1)+1)/2$ and $b_2=(N(A+1)+2)/2$. Then Eq.~\eqref{apphg1f2diff} is 
\begin{equation}
\epsilon^2 z^2\frac{d^3}{dz^3} y + \left[(A+1) z \epsilon+\frac{5}{2} z \epsilon^2\right]\frac{d^2}{dz^2} y +\left[\frac{(A+1)^2}{4}+\frac{3(A+1)}{4}\epsilon+(\frac{1}{2}-z)\epsilon^2 \right]\frac{d}{dz}y+\epsilon y=0,
    \label{apphg1F2diff01}
\end{equation}
 where $\epsilon=1/N$ and $y={{}_1 F_2}(-N;(N(A+1)+1)/2,(N(A+1)+2)/2;z)$. In a similar fashion as in Appendix~\ref{appendix:hqfp} we search for a solution in the $N\to\infty$ limit and use a perturbation series representation for small $\epsilon$, i.e. $y(z)=\sum_{k=0}^\infty y_k(z)$. We use the boundary conditions $y_(0)=1$ and $\forall k\geq 1$ $y_k(0)=0$. Introduction of this representation into Eq.~\eqref{apphg1F2diff01} yields the solution for $y_0$
 \begin{equation}
 y_0(z)=1
     \label{apphg2F1y0}
 \end{equation}
The equation for $y_1(z)$ is 
\begin{equation}
\frac{(A+1)^2}{4}\frac{dy_1}{dz}+ 
(A+1)z\frac{d^2y_0}{dz^2}+\frac{3(A+1)}{4}\frac{dy_0}{dz}+y_0=0
    \label{apphg1F2y1}
\end{equation}
that together with Eq.~\eqref{apphg2F1y0} and the boundary conditions leads to
\begin{equation}
y_1(z) = -\frac{4}{(A+1)^2}z.
    \label{apphg1F2y1s}
\end{equation}
For any $k\geq 2$ we can write the recursive  equation for $y_k$
\begin{equation}
\begin{array}{ll}
\displaystyle
y_k(z)= &
\\
-\frac{4}{(A+1)^2}\int \left\{
\left[z^2\frac{d^3y_{k-2}}{dz^3}+\frac{5z}{2}\frac{d^2y_{k-2}}{dz^2}+(\frac{1}{2}-z)\frac{dy_{k-2}}{dz} \right]+\left[(A+1)z\frac{d^2 y_{k-1}}{dz^2}+\frac{3(A+1)}{4}\frac{dy_{k-1}}{dz}+y_{k-1} \right]
\right\}\,dz &
\end{array}
    \label{apphg1F2yk}
\end{equation}
Inspection of Eq.~\eqref{apphg1F2yk} shows that $y_k(z)$ is a polynomial of degree $k$ without a constant term. According to Eq.~\eqref{apphg1F2yk} the leading term of $y_k(z)$ is $(-4)^k z^k\Big/(A+1)^{2k}k!$. Since ${{}_1 F_2}(-N;(N(A+1)+1)/2,(N(A+1)+2)/2;z) = \sum_{k=0}^\infty y_k(z)/N^k$, the hypergeometric function ${}_1 F_2$ is a polynomial in $z$. The highest ratio of the power of $z$ and the power of $N$ is $1$. 
This fact will become important in the main text when addressing the convergence of $Q_t(N)$ to its limit form.
Specifically, when taking into account only the terms that are powers of $z/N$ in the expansion $\sum_{k=0}^\infty y_k(z)/N^k$  and using the fact that $\exp(-z)=\sum_{k=0}^\infty (-1)^k z^{k}/k!$ we obtain that
\begin{equation}
\displaystyle
{{}_1 F_2}(-N;(N(A+1)+1)/2,(N(A+1)+2)/2;z)\underset{N\to\infty}{\sim}
e^{-\frac{4}{(A+1)^2}\frac{z}{N}}.
    \label{apphg1F2Nlarge}
\end{equation}

\bibliographystyle{apsrev4-1} 
\bibliography{bibnonGaussian} 

\end{document}